# The Role of Asymmetries in Thermal Non-Equilibrium

James A. Klimchuk

Heliophysics Division
NASA Goddard Space Flight Center
8800 Greenbelt Road
Greenbelt, MD  20771  USA
James.A.Klimchuk@nasa.gov

Manuel Luna

Instituto de Astrofisica de Canarias
E-38200 La Laguna
Tenerife, Spain
Departamento de Astrofisica
Universidad de La Laguna
E-38206 La Laguna
Tenerife, Spain

**Abstract**

Thermal non-equilibrium (TNE) is a fascinating situation that occurs in coronal magnetic flux tubes (loops) for which no solution to the steady-state fluid equations exists. The plasma is constantly evolving even though the heating that produces the hot temperatures does not. This is a promising explanation for isolated phenomena such as prominences, coronal rain, and long-period pulsating loops, but it may also have much broader relevance. As known for some time, TNE requires that the heating be both (quasi) steady and concentrated at low coronal altitudes. Recent studies indicate that asymmetries are also important, with large enough asymmetries in the heating and/or cross-sectional area resulting in steady flow rather than TNE. Using reasonable approximations, we have derived two formulae for quantifying the conditions necessary for TNE. As a rough rule of thumb, the ratio of apex to footpoint heating rates must be less than about 0.1, and asymmetries must be less than about a factor of 3. The precise values are case dependent. We have tested our formulae with 1D hydrodynamic loop simulations and find a very acceptable agreement. These results are important for developing physical insight about TNE and assessing how widespread it may be on the Sun.

## 1. Introduction

Thermal non-equilibrium (TNE) is one of the most fascinating phenomena in solar physics, in part because it is so counter-intuitive. In the presence of steady heating, coronal loops are normally in an equilibrium state---either a static equilibrium or an equilibrium with steady end-to-end flow. However, under certain conditions no equilibrium exists. The loop is inherently dynamic, evolving continuously even though the heating is unchanging. This is known as thermal non-equilibrium (Antiochos & Klimchuk 1991). It occurs when the heating decreases strongly with altitude and, as we will show, asymmetries are not too great.

Under TNE conditions, the loop atmosphere is constantly adjusting to imbalances in the energies and forces, essentially searching for a nonexistent equilibrium. It periodically enters a phase of catastrophic cooling, usually but not always leading to the formation of a cold, high density condensation in the corona. The condensation slides down the loop leg, and the process repeats.

It is sometimes said that TNE results from thermal instability, but this is a misconception. Instability requires that there be an equilibrium to go unstable in the presence of a perturbation (Parker 1994, pg. 1023). As the name implies, TNE means that no equilibrium exists, stable or otherwise. It is preferable to use the phrase "thermal runaway" rather than "thermal instability" when describing the plasma evolution that occurs under TNE conditions. The distinction between TNE and thermal instability is further discussed in Klimchuk (2019), where we also consider situations where the two may be interconnected.

Thermal non-equilibrium is believed to play an important role in explaining a number of phenomena. A prime example is "coronal rain" seen in H$_\alpha$ and other cool emissions (Schrijver 2001; Muller, Peter, & Hansteen 2004; Antolin, Shibata, & Vissers 2010). If the magnetic field has a dip---a region of upward concavity---the cold condensation can settle in the dip and grow in size. This has become the standard explanation for prominence formation (Antiochos et al. 1999; Karpen, Antiochos, & Klimchuk 2006; Luna, Karpen, & DeVore 2012).

Recently, another phenomenon has also been attributed to TNE. Extreme ultra-violet (EUV) observations reveal that active region loops sometimes exhibit long-period (several-hour) pulsations, with multiple cycles spanning well over a day (Auchere et al. 2014; Froment et al. 2015, 2017). The behavior is consistent with TNE. TNE may also explain periodic density enhancements observed in the solar wind, the idea being that it triggers quasi-regular reconnection episodes at the tips of helmet streamers (Viall & Vourlidas 2015; Antiochos et al. 2018). These latter two phenomena involve multiple repeating TNE cycles, requiring that the heating conditions be essentially unchanging for long periods of time. For coronal rain and prominences, the quasi-steady conditions need apply only long enough for one cycle to occur.

It is not known whether TNE is a common, widespread occurrence or a relative oddity. Although coronal rain is regularly observed (Schrijver 2001; Antolin & Rouppe van der Voort 2012), it seems that only a small fraction of the magnetic flux tubes that make up the corona actually participate (a rigorous estimate has yet to be made). On the other hand, active region models based on parameterized turbulent wave heating seem to exhibit a large amount of TNE behavior (Mok et al. 2016). Klimchuk, Karpen, & Antiochos (2010) investigated the possibility that ordinary coronal loops at warm (~ 1 MK) temperatures might be explained by TNE. Such loops have significantly higher densities than expected for static equilibrium, and TNE seemed a promising explanation. However, it was found that other observed properties of the loops are violated if a cold condensation is formed. This is true even if loops are composed of multiple sub-resolution strands that are behaving independently such that no individual condensation can be detected.

Mikic et al. (2013) subsequently showed that the assumptions of symmetric heating and uniform cross sectional area in the Klimchuk et al. (2010) models restrict the range of possible behavior. By relaxing those assumptions, they found a new type of TNE in which the plasma never cools below 1 MK. This was termed "incomplete condensation." Whether solutions of this type are consistent with all the observed properties of coronal loops has yet to be established (Lionello et al. 2013; Winebarger et al. 2014, 2016, 2018).

It is important to remember that most of the emission from active regions is contained in a diffuse component, not the distinct loops that the eye is drawn to in images. Loops are typically only small (10-30%) enhancements over the background (Viall & Klimchuk 2011). TNE could be playing an important role in the diffuse corona, even if it turns out to not be a generic explanation for distinct loops (Downs et al. 2016; Winebarger et al. 2016; Yung et al. 2016).

To better understand how common TNE may be, Froment et al. (2018) recently performed a series of loop simulations to determine the dependence on model parameters. They found that TNE is quite sensitive to the parameters, with the conditions for TNE being rather restrictive. Asymmetries are very important, as also found by others (e.g., Mikic et al. 2013).

The purpose of our investigation reported here is to better understand the physical causes of TNE and the role played by asymmetries. Asymmetries in both heating and cross-sectional area are considered. Using reasonable approximations and making an innovative use of equilibrium loop scaling laws, we derive two formulae for predicting whether loops should be in static equilibrium, steady flow equilibrium, or TNE. We have tested the formulae with numerical simulations and find good agreement overall. The formulae and simulations are presented in Sections 3 and 4, but first we discuss the basic physical principles underlying thermal non-equilibrium.

## 2. Basic Concepts

The physics of loops in static equilibrium has been well understood for many years (e.g., Rosner, Tucker, & Vaiana 1978; Craig, McClymont, & Underwood 1978; Vesecky, Antiochos, & Underwood 1979). There is a balance between the input energy (coronal heating), thermal conduction, and radiation. Roughly one-third of the energy deposited in the coronal part of the loop is radiated directly to space. The other two-thirds is thermally conducted down the legs to the transition region and radiated from there. Although the transition region is very thin, its emissivity---radiative loss rate per unit volume---far exceeds that of the corona. The energy balance is indicated schematically on the left of Figure 1 for the case of coronal heating that is uniform along the loop. The lengths of the arrows represent the magnitudes of the terms in the energy equation. Blue arrows show the cooling rate (erg cm$^{-3}$ s$^{-1}$) from thermal conduction, which is the divergence of the thermal conduction flux. It is not the flux itself, which increases steadily downward from the apex to the transition region, and then decrease dramatically through the transition region to zero at the top of the chromosphere. The conductive cooling rate arrows

6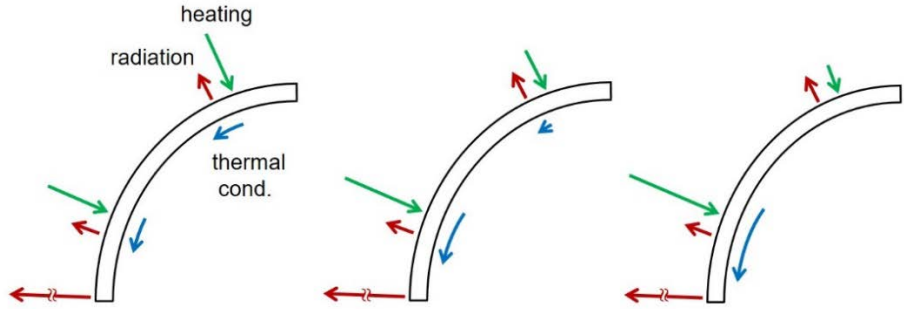

**Figure 1.** Schematic representation of the energy balance in three symmetric coronal loops with uniform coronal heating (left), modest concentration in the lower corona (middle), and strong concentration in the lower corona (right). Arrows indicate the terms in the energy equation (erg cm$^{-3}$ s$^{-1}$) associated with heating (green), radiation (red), and thermal conduction (blue). A large energy input from thermal conduction in the transition region is not shown.

are drawn parallel to the loop axis to indicate that the energy is not lost to the loop, but rather transported downward.

The two groups of three arrows in the upper and lower corona indicate a local energy balance, with the green arrow (heating) having a length equal to the combined length of the red (radiative cooling) and blue (conductive cooling) arrows. As already mentioned, the radiative cooling rate is much greater in the transition region than in the corona, so the red arrow at the bottom should actually be much longer than shown. For simplicity, we chose not to add a similarly long blue arrow representing the heating from thermal conduction in the transition region. Only the left side of the loop is shown due to symmetry.

The temperature profile in the loop (temperature as a function of position along the loop axis, typically denoted by curvilinear coordinate *s*), has the characteristic shape shown on the right side of the top row in Figure 2. Temperature rises steeply in the transition region at both ends and levels off in the corona. The profile is modestly rounded.

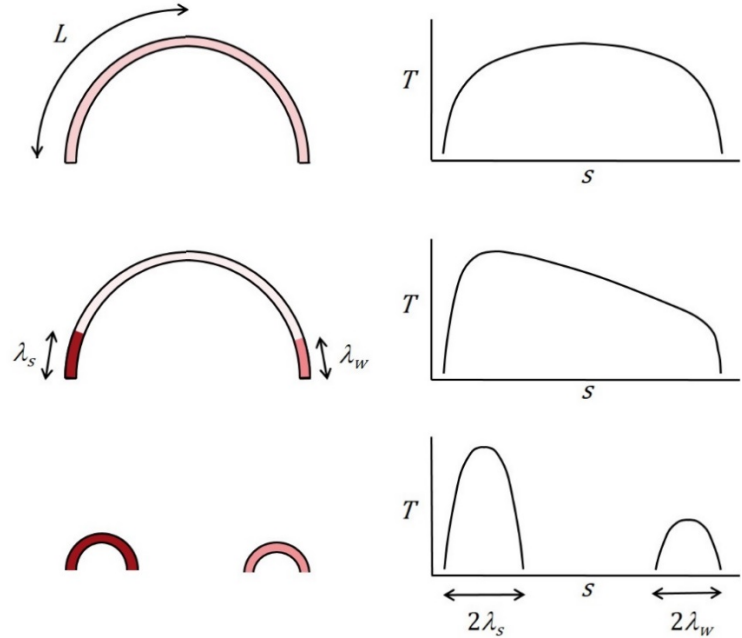

**Figure 2.** Schematic representation of a symmetric loop with uniform heating (top) and an asymmetric loop with heating concentrated in the lower legs (middle). Coloring indicates heating rate: red-strong, pink-weak, rose-intermediate. Right shows temperature versus position along the loop axis. Bottom row shows two mini-loops that are representative of the conditions in the lower legs of the asymmetric full loop.

Suppose that some of the heating that was deposited in the upper corona is instead deposited in the lower corona, with the total heating remaining unchanged. To achieve equilibrium, the combined cooling from radiation and thermal conduction must decrease in the upper part. This is accomplished primarily by a decrease in conduction. There is also some reduction in radiation, but the effect is smaller for the following reason. Thermal conduction redistributes energy within the loop, but only radiation can remove it. Radiation has a strong dependence on density, varying as $n^2$. Because there is minimal pressure stratification in most loops, density cannot decrease substantially in the upper part without also decreasing substantially everywhere, including the transition region. But this is not allowed since the total radiative losses must be unchanged in order to balance the unchanged total heating. The reduction in cooling in the upper corona is therefore accomplished primarily by a reduction in thermal conduction losses.

The modified energy balance is represented by the middle sketch in Figure 1. The green (heating) and blue (conductive cooling) arrows are shorter in the upper corona and longer in the lower corona compared to uniform heating. The red (radiative cooling) arrows are largely unchanged. To modify the conduction cooling as required, the temperature profile must flatten in the central part of the loop and steepen slightly at the ends. It is less rounded than the uniform heating case.

Now suppose that we transfer even more heating from the upper to lower corona, again keeping the total unchanged. Conduction cooling must decrease further at the top, and the temperature profile flattens even more. There is a limit to how far this can proceed. Once the temperature gradient reaches zero at the apex and the profile becomes perfectly flat, no further decrease in conduction cooling is possible. This state is represented on the right side of Figure 1. If still more heating is transferred from the upper to lower corona, an equilibrium is no longer possible. There is an unbalanced excess in radiation.[1] This is the essence of thermal non-equilibrium. A basic understanding of the cause of TNE can be traced back to Serio et al. (1981), although the ensuing time-dependent behavior was not known at the time.

We have thus far considered loops that are perfectly symmetric. This is not, of course, a good representation of reality. Any asymmetry in the heating rate or cross sectional area will produce an end-to-end flow (e.g., Mariska & Boris 1983; Craig & McClymont 1986). The flow carries an energy flux---primarily enthalpy if the flow is subsonic---and a gradient in this flux will provide heating or cooling. Is it possible for the heating associated with a steady flow to balance the excess radiation in the upper part of a loop that would otherwise experience TNE? This is indicated schematically in Figure 3. The answer to the question is yes, as we now show.

---

[1] In principle, a temperature dip at the apex can provide a small amount of thermal conduction heating to balance the excess radiation (Aschwanden, Schrijver, & Alexander 2001; Winebarger, Warren, & Mariska 2003), but equilibria of this type appear to be unstable (Winebarger et al. 2003; Martens 2010).

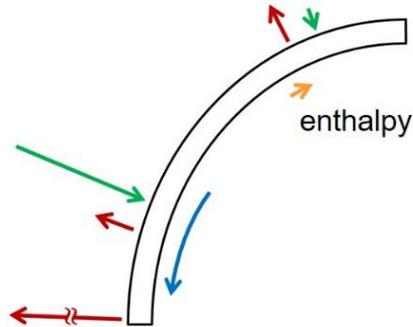

**Figure 3.** Schematic representation of the energy balance in an asymmetric loop with steady end-to-end flow from the left to right. Arrows indicate the terms in the energy equation, as in Figure 1. The yellow arrow represents heating from a divergence of the enthalpy flux.

### 3. Predictive Formulae

Symmetric loops with uniform heating are in static equilibrium and obey well-known scaling laws that relate any three of the fundamental loop parameters: length, heating rate, pressure (roughly constant along most loops), apex temperature, and apex density (e.g., Rosner et al. 1978; Craig et al. 1978; Martens 2010). Perhaps the most famous of these is $T_a \propto (PL)^{1/3}$, which applies for a particular radiative loss function. The scaling laws are also valid with mildly non-uniform heating. In that case the characteristic, or spatially-averaged, heating rate is used. We now derive new versions of the scaling laws that take into account non-uniform cross-sectional area.

Consider a loop that is symmetric, so that only half need be treated. The heating can be mildly non-uniform. A static equilibrium is established in which the total energy input is balanced by the combined radiative losses from the corona and transition region:

$$A_c L Q \approx A_c R_c + A_{tr} R_{tr} \ .$$

Eq. 1

Here, $Q$ is the spatially averaged volumetric heating rate (erg cm$^{-3}$ s$^{-1}$), $L$ is the coronal half-length (distance from the top of the transition region to the apex), $R_c$ and $R_{tr}$ are the radiative losses per unit area (erg cm$^{-2}$ s$^{-1}$) integrated along the coronal and transition region sections, respectively, and $A_c$ and $A_{tr}$ are the cross-sectional areas averaged along the coronal and transition region sections. Following Klimchuk, Patsourakos, and Cargill (2008), the boundary between the transition region and corona is defined to be the place where thermal conduction switches from being a heating term above to a cooling term below. $R_c$ is defined as

$$R_c \equiv \frac{1}{A_c} \int_{cor} A(s) \, \varepsilon(s) \, ds \ ,$$

Eq. 2

where $\varepsilon(s)$ is the emissivity and the integral is taken over the coronal section. A similar expression applies for the transition region radiative losses.

The left hand side of Equation (1) assumes that the transition region is so thin that any "coronal heating" energy deposited there is ignorable compared to that in the corona (accounted for with the approximately equals sign). Losses from radiation are large, on the other hand, because of the strong emissivity, and these losses are powered by the intense downward thermal conduction flux from the corona:

$$A_0 F_0 \approx A_{tr} R_{tr} \ ,$$

Eq. 3

where $F_0$ and $A_0$ are the conduction flux and area at the top of the transition region. Noting that $T^{5/2} dT/ds = (2/7) d/ds(T^{7/2})$, we can approximate the conduction flux as

$$F_0 \approx \frac{2}{7} \kappa_0 \frac{T_a^{7/2}}{L},$$

Eq. 4

where $T_a$ is the temperature at the apex and $\kappa_0$ is the coefficient of thermal conduction.

Combining Equations (1), (3), and (4), we get

$$T_a \approx \left( \frac{7}{2\kappa_0} QL^2 \frac{A_{tr}/A_0}{1/c_1 + A_{tr}/A_c} \right)^{2/7},$$

Eq. 5

where

$$c_1 \equiv \frac{R_{tr}}{R_c}$$

Eq. 6

is a constant introduced by Klimchuk et al. (2008) and discussed at length by Cargill, Bradshaw, & Klimchuk (2012). The ratio of transition region to coronal radiative losses is not actually a constant, as there is some dependence on loop length and apex temperature under realistic conditions (with gravitational stratification and a non-trivial radiative loss function). However, models of equilibrium loops with uniform cross section indicate that a value of 2 is a reasonable approximation under most circumstances. We note that Equation (5) is in good agreement with the expanding cross section analytical solutions of Martens (2010) when $c_1 = 2$ and the loss function has a constant power law slope of -0.5.

If we assume that the loop expands only minimally in the thin transition region, so that $A_{tr} \approx A_0$, and define an expansion factor

$$\Gamma \equiv \frac{A_c}{A_{tr}} ,$$

Eq. 7

we can rewrite Equation (5) as

$$T_a \approx \left( \frac{7}{2\kappa_0} Q L^2 \frac{c_1 \Gamma}{c_1 + \Gamma} \right)^{2/7} .$$

Eq. 8

This is our new scaling law for temperature. We also seek a scaling law for pressure.

Temperature and density are uniform to within 50% along the coronal section of a symmetric equilibrium loop (Klimchuk et al. 2008; Cargill et al. 2012), so the radiative losses can be approximated by

$$R_c \approx n^2 \Lambda_0 T_a^b L ,$$

Eq. 9

where $n$ is the average electron number density and $\Lambda(T) = \Lambda_0 T^b$ is a simplified form of the optically-thin radiative loss function. Combining the above equations and using the ideal gas law for a fully ionized hydrogen plasma,

$$P = 2knT ,$$

Eq. 10

where $k$ is Boltzmann's constant, we obtain the following expression for the total (electron plus proton) pressure:

$$P \approx \left( \frac{8}{7} \frac{\kappa_0 k^2}{\Lambda_0} \frac{1}{c_1} \frac{A_0}{A_{tr}} \right)^{\frac{1}{2}} \frac{T_a^{\frac{11-2b}{4}}}{L} .$$

Eq. 11

This is the average pressure along the loop, but it is a reasonable approximation to the pressure at all locations, with the exception of especially tall and/or cool loops where there is significant gravitational stratification. Equations (8) and (11) give the apex temperature and pressure in terms of "known" quantities ($Q$, $L$, $\Gamma$, $A_0/A_{tr} \approx 1$). We now use them to derive the two conditions necessary for TNE.

### 3.1. Heating Sufficiently Concentrated at Low Altitudes

As we have discussed, TNE requires that the energy input to the upper corona be too small to balance the local radiative losses, a situation that occurs when the heating decreases rapidly with height. Our approach is to determine the most extreme conditions under which equilibrium is possible, as this sets the threshold for TNE. With heating that is strongly height dependent, the equilibrium temperature profile has two distinctive knees where a steep rise in the lower legs rolls over rather rapidly into a long flat section. This is shown schematically in the sketch at the middle-right of Figure 2. The heating in this example is asymmetric, so the flat central section is inclined. With symmetric heating it is horizontal. The top of the knee is located approximately one heating scale length, $\lambda$, above the transition region on each side.

Since most of the energy input to the loop is deposited in the lower leg, equilibrium conditions will be determined largely by the need to satisfy energy balance there. We can therefore think of the lower leg, extending from the coronal footpoint to the top of the knee, as being one-half of a small symmetric loop of total length $2\lambda$, as shown at the bottom of the figure. The conditions in the lower leg of the full loop will be similar to those in the imaginary mini-loop. At this stage, we are considering a full loop that is symmetric, so the two mini-loops on either side are identical. We will later relax this assumption. We apply Equations (8) and (11) to the mini loops, with $L$ replaced by $\lambda$, $T_a$ replaced by $T_\lambda$, the temperature at location $\lambda$ above the footpoint, $Q$ replaced by $Q_\lambda$, the average heating rate over the first scale length, and $\Gamma$ defined in terms of coronal area $A_\lambda$ that is the average over the first heating scale length. Note that $\lambda$ refers to distance along the loop axis, which is different from vertical height.

Since the temperature of the actual loop (of length $2L$) is roughly constant beyond the knee when the heating is symmetric, $T_\lambda$ will be the approximate temperature throughout the entire upper section. The pressure is also nearly uniform and equal to the pressure of the mini-loop, $P_\lambda$, given by the modified version of Equation (11). Uniform temperature and pressure imply uniform density, so we can express the radiative loss rate in the upper section as

$$n_\lambda^2 \Lambda_0 T_\lambda^b \approx \left(\frac{P_\lambda}{2kT_\lambda}\right)^2 \Lambda_0 T_\lambda^b .$$

Eq. 12

TNE occurs when this radiation rate exceeds the local heating rate, which has a minimum value $Q_{min}$ at the loop apex. Substituting for $T_\lambda$ and $P_\lambda$, making use of the ideal gas law, and assuming $A_{tr} \approx A_0$, we obtain the first condition for TNE:

$$\frac{Q_{min}}{Q_\lambda} < \left(1 + \frac{c_1}{\Gamma_\lambda}\right)^{-1} .$$

Eq. 13

There is a threshold for how small the heating must be at the top of the loop compared to the base in order for TNE to occur. This threshold depends on the expansion factor and is less severe for greater expansion. In other words, loops with large expansion are more prone to TNE.

Although Equation (13) was derived for symmetric loops, it should also apply to mildly asymmetric loops. In those situations, we use $Q_\lambda$ and $\Gamma_\lambda$ for the strongly heated leg when applying the formula.

### 3.2. Sufficiently Small Asymmetries

The above discussion assumes a symmetric loop, where the two possible states are TNE and static equilibrium. In general, loops are not symmetric, either in their heating or their cross-sectional area. This leads to a pressure imbalance in the two legs, which drives an end-to-end flow along the loop axis. If the flow is not too fast, the conditions in the two legs can once again be approximated by two imaginary mini-loops in static equilibrium. Unlike the symmetric case, the mini-loops now have different lengths, heating rates, "apex" temperatures, and/or area expansion factors. Equations (8) and (11) become

$$T_{s,w} \approx \left(\frac{7}{2\kappa_0} Q_{s,w} \lambda_{s,w}^2 \frac{c_1 \Gamma_{s,w}}{c_1 + \Gamma_{s,w}}\right)^{2/7}$$

Eq. 14

$$P_{s,w} \approx \left(\frac{8}{7}\frac{\kappa_0 k^2}{\Lambda_0}\frac{1}{c_1}\right)^{1/2} \frac{T_{s,w}^{(11-2b)/4}}{\lambda_{s,w}},$$

Eq. 15

where the subscripts "s" and "w" distinguish between the strongly heated and weakly heated legs. As discussed earlier, the asymmetry-induced flow carries an energy flux. Because heating, thermal conduction, and radiation are much weaker in the upper section of the full loop than in the lower legs, the flow may be energetically important there even when it plays a small role in the lower legs. If the gradient in the energy flux is sufficiently large, the associated heating may be enough to balance the excess radiation relative to $Q_{min}$ and prevent TNE from occurring.[2] A steady state equilibrium would result instead.

For subsonic flow, we can ignore kinetic energy compared to enthalpy. The validity of this approximation is discussed later. We also ignore the effect of gravity, since we are concerned with the upper part of the loop, where the magnetic field is likely to be substantially horizontal. From the steady state energy equation (e.g., Karpen et al. 2005), we then obtain a simplified condition for energy balance in the upper loop:

$$-\frac{1}{A}\frac{d}{ds}\left(\frac{5}{2}PvA\right) \geq n^2\Lambda_0 T^b - Q_{min} ,$$

Eq. 16

where $v$ is the field-aligned velocity, and we have taken the ratio of specific heats to be $\gamma = 5/3$. We write this as an inequality because we are interested in the conditions that must be exceeded in order for TNE to be avoided. (Our final predictive formula gives the limiting conditions where TNE is expected.) Since the mass flux is constant along a loop in steady state, we can write the left hand side of Equation (16) as

$$-\frac{1}{A}\frac{d}{ds}\left(\frac{5}{2}PvA\right) = -\frac{5kJ}{A}\frac{dT}{ds} ,$$

Eq. 17

where $J = nvA$ is the electron mass flux.

Our scaling laws allow us to express temperature, pressure, and density in terms of known quantities, but we need to eliminate velocity (mass flux). For this we turn to the steady state momentum equation. Again we ignore gravity, leaving

$$\frac{dP}{ds} \approx -\frac{1}{A}\frac{d}{ds}(A\rho v^2) ,$$

---

[2] Note that $Q_{min}$ will be spatially offset from the apex if the heating is asymmetric.

Eq. 18

which simplifies to

$$\frac{dP}{ds} \approx -\frac{d}{ds}(\rho v^2)$$

Eq. 19

if cross-sectional area varies more weakly with position than does kinetic energy. Integrating between the strongly and weakly heated ends and assuming that the flow is appreciably slower at the strongly heated end, where the mini-loop predicts higher density, we obtain

$$P_s - P_w \approx \frac{m_p J^2}{n_w A_w^2} \ .$$

Eq. 20

Making use of the ideal gas law, this can be rewritten as

$$\frac{J}{A} \approx \left(\frac{2k}{m_p}\right)^{1/2} \xi n_s T_s^{1/2} \ ,$$

Eq. 21

where

$$\xi \equiv \left(1 - \frac{P_w}{P_s}\right)^{1/2} \frac{A_w}{A} \left(\frac{n_w}{n_s}\right)^{1/2} \ .$$

Eq. 22

Since we intend to substitute for $J/A$ in Equation (17), $A$ is the area at the location of $Q_{min}$, where enthalpy heating is most critical.

From Equations (14) and (15),

$$\frac{P_w}{P_s} \approx \left[\frac{Q_w}{Q_s} \frac{\Gamma_w}{\Gamma_s} \frac{\Gamma_s + c_1}{\Gamma_w + c_1}\right]^{\frac{11-2b}{14}} \left(\frac{\lambda_w}{\lambda_s}\right)^{\frac{4-2b}{7}} \ .$$

Eq. 23

The choice $b = -0.5$ provides a single power-law radiative loss function that is a reasonable representation of the actual loss function over a wide range of coronal temperatures (Rosner et al. 1978). The pressure ratio then becomes

$$\frac{P_w}{P_s} \approx \left[\frac{Q_w}{Q_s} \frac{\Gamma_w}{\Gamma_s} \frac{\Gamma_s + c_1}{\Gamma_w + c_1}\right]^{\frac{6}{7}} \left(\frac{\lambda_w}{\lambda_s}\right)^{\frac{5}{7}}.$$

Eq. 24

It varies approximately linearly with the heating and area asymmetry ratios. As discussed later, our simulations show that the transition between TNE and steady flow behavior occurs when the asymmetry is roughly a factor of 3. We therefore approximate $P_w/P_s$ as 1/3 in Equation (22). Substituting for density using Equations (14) and (15) and the idea gas law, we have

$$\xi = \left(\frac{2}{3}\right)^{1/2} \frac{A_w}{A} \left[\frac{Q_w}{Q_s} \frac{\Gamma_w}{\Gamma_s} \frac{\Gamma_s + c_1}{\Gamma_w + c_1}\right]^{\frac{7-2b}{28}} \left(\frac{\lambda_w}{\lambda_s}\right)^{-\frac{b}{7}}.$$

Eq. 25

Our final approximation is that the temperature gradient in the upper part of the loop can be expressed as

$$\frac{dT}{ds} \approx \frac{T_s - T_w}{2L}.$$

Eq. 26

Combining Equations (14), (15), (16), (17), (20), (25), and (26), and using $n_s$ and $T_s$ to evaluate the radiation term in Equation (16), we obtain the second condition that must be satisfied for TNE to occur:

$$\left(\frac{Q_s}{Q_w}\right)\left(\frac{\lambda_s}{\lambda_w}\right)^2 \left(\frac{\Gamma_s}{\Gamma_w}\right)\left(\frac{\Gamma_w + c_1}{\Gamma_s + c_1}\right) < \left[1 - \eta(1-\delta)\frac{L}{\lambda_s}\left(\frac{7}{2\kappa_0}Q_s\lambda_s^2 \frac{c_1 \Gamma_s}{\Gamma_s + c_1}\right)^{(1+2b)/14}\right]^{-7/2},$$

Eq. 27

where

$$\eta \equiv \left(\frac{4}{175}\frac{\kappa_0 \Lambda_0 m_p}{k^3 c_1}\right)^{1/2}\frac{1}{\xi}$$

Eq. 28

and

$$\delta = \left(1 + \frac{c_1}{\Gamma_s}\right)\left(\frac{Q_{min}}{Q_s}\right).$$

Eq. 29

Things simplify considerably when $b = -0.5$, in which case Equation (27) becomes

$$\left(\frac{Q_s}{Q_w}\right)\left(\frac{\lambda_s}{\lambda_w}\right)^2\left(\frac{\Gamma_s}{\Gamma_w}\right)\left(\frac{\Gamma_w + c_1}{\Gamma_s + c_1}\right) < \left[1 - \eta(1-\delta)\frac{L}{\lambda_s}\right]^{-7/2},$$

Eq. 30

and Equation (25) becomes

$$\xi = \left(\frac{2}{3}\right)^{1/2}\frac{A_w}{A}\left[\frac{Q_w}{Q_s}\frac{\Gamma_w}{\Gamma_s}\frac{\Gamma_s + c_1}{\Gamma_w + c_1}\right]^{2/7}\left(\frac{\lambda_w}{\lambda_s}\right)^{-1/14}.$$

Eq. 31

The dependence of $\xi$ on the asymmetry ratios is very weak, so we approximate these terms as unity:

$$\xi \approx \left(\frac{2}{3}\right)^{1/2}\frac{A_w}{A}.$$

Eq. 32

With the radiative loss function

$$\Lambda(T) = \Lambda_0 T^b = 10^{-18.75}T^{-1/2}$$

Eq. 33

and $c_1 = 2$, Equation (30) becomes

$$\left(\frac{Q_s}{Q_w}\right)\left(\frac{\lambda_s}{\lambda_w}\right)^2 \left(\frac{\Gamma_s}{\Gamma_w}\right)\left(\frac{\Gamma_w + 2}{\Gamma_s + 2}\right) < \left[1 - 4.3\text{x}10^{-2}\frac{A}{A_w}(1-\delta)\frac{L}{\lambda_s}\right]^{-7/2},$$

Eq. 34

where $A$ and $A_w$ are the cross sectional areas at, respectively, the location of minimum heating and one heating scale length above the transition region on the weakly heated side. Since the location of minimum heating is displaced from the apex toward the weakly heating footpoint, the area ratio is not likely to be large. It ranges between 1.0 and 2.2 in our simulations.

TNE is predicted to occur if the heating is sufficiently concentrated at low altitudes, as given by Equation (13), and if the asymmetry in heating and/or cross-sectional area is not too great, as given by Equation (34). Both inequalities must be satisfied. Note that the form of the right hand side of Equation (34) is reasonable. The larger it is, the more likely TNE is to occur, and it is larger when $\lambda_s/L$ and $Q_{min}/Q_s$ are smaller, both of which promote TNE according to Equation (13).

Our derivations involved a number of approximations and simplifications, so it is important that the final formulae be validated with rigorous numerical simulations, as we now discuss.

## 4. Numerical Simulations

The heating rate and coronal cross-sectional area in the formulae are defined to be averages over the first heating scale length above the transition region, while the transition region cross-sectional area is the average over the transition region. It is easier to identify the top of the chromosphere in numerical simulations than it is to identify the top of the transition region, so we redefine some of the parameters: $Q$ is now the heating rate at the top of the chromosphere; $\lambda$ is the first heating scale length above it; $A_{tr}$ is the cross-sectional area at the top of the chromosphere; $A_c$ is the average cross-sectional area one heating scale length above it, and $L$ is the loop half-length from the top of the chromosphere to the apex. We refer to $Q$ and $A_{tr}$ as the footpoint heating rate and footpoint area, respectively. Subscripts "s" and "w" are left off for simplicity. These new definitions should not greatly affect the predictions as long as the transition region is thin compared to $\lambda$. However, caution should be exercised when comparing with numerical simulations that use techniques to artificially broaden the transition region to make it better resolved (e.g., Lionello et al. 2009; Johnston & Bradshaw 2019). We do not employ those techniques here.

We use an exponential heating function in our simulations, in which case the average heating rate over one scale length is a factor of 0.63 smaller than the bottom value. Taking this into account, as well as the finite thickness of the transition region, we introduce a factor $\alpha = 0.45$ that relates the average heating rate one scale length above the transition region, $Q_\lambda$, to the footpoint heating rate, $Q$, at the top of the chromosphere:

$$Q_\lambda = \alpha Q \ .$$

Eq. 35

Different factors apply to non-exponential heating functions. Our first condition for TNE then becomes

$$\frac{Q_{min}}{Q_s} < \alpha \left(1 + \frac{c_1}{\Gamma_\lambda}\right)^{-1} ,$$

Eq. 36

and $\delta$ in the second condition becomes

$$\delta = \left(1 + \frac{c_1}{\Gamma_s}\right)\left(\frac{Q_{min}}{\alpha Q_s}\right) ,$$

Eq. 37

where $Q_s$ is the footpoint heating rate on the strongly heated side. $A_w$ in the second condition, Equation (34), is taken to be the cross sectional area one heating scale length above the chromosphere on the weakly heated side. $A$ is still the area at the location of $Q_{min}$.

We perform 1D hydrodynamic simulations using the Adaptively Refined Godunov Solver (ARGOS) (Antiochos et al. 1999). To facilitate a direct comparison with our formulae, we use the simplified radiative loss function given by Equation (33), modified to have a $T^3$ dependence in the lower transition region ($T < 10^5$ K) and to fall steeply to zero over a 500 K temperature interval at the top of the chromosphere:

$$\Lambda(T) = \begin{cases} 10^{-18.75} \times T^{-1/2} & T \geq 10^5 \\ 10^{-36.25} \times T^3 & 3\times10^4 \leq T < 10^5 \\ 10^{-25.518} \times (T - 2.95\times10^4) & 2.95\times10^4 \leq T < 3\times10^4 \end{cases}.$$

Eq. 38

This produces a nearly isothermal chromosphere at $3\times10^4$ K. The loop lies in a vertical plane and has a semi-circular shape of half-length $L = 40$ Mm. Added to each end are 40 Mm long chromospheric sections (or 45 Mm for some cases), providing a large reservoir of mass and assuring that the closed boundaries at the ends of the model do not influence the evolution.

We note that actual radiative loss function is considerably more complex, especially at lower temperatures. This will affect the rate of final thermal collapse of a forming condensation, but it is unlikely to have a big impact on whether a given set of parameters leads to equilibrium or TNE behavior. We discuss this further below.

An important aspect of time-dependent simulations that must be taken into account is the displacement of the chromosphere. As the coronal pressure evolves, the top of the chromosphere moves up and down. This is a never ending process during TNE, and it can also be important when a loop settles into a true equilibrium after starting from some approximate initial state. It is crucial that the parameters in the formulae be evaluated at an appropriate time. Values based on the initial location of the chromosphere can lead to erroneous predictions.

We begin all of our simulations with an approximate static equilibrium solution to a specified uniform heating. The loop is allowed to settle into the true equilibrium, at which point we gradually transition the heating to the exponential form we wish to investigate. The simulation is then run for many hours of solar time, leading either to a final static or steady flow equilibrium or to a series of TNE cycles. All formula parameters are evaluated after the new equilibrium has been established or, in the case of TNE, at a time during one of the later cycles when the temperature profile is flat (constant slope), just before a dip forms leading to a thermal runaway.

Our heating profile decreases exponentially with distance from both ends of the loop:

$$Q(s) = Q_{s,i} e^{-(s-s_{ch})/\lambda_s} + Q_{w,i} e^{-(s_{ch}+2L-s)/\lambda_w} + Q_{bkg} ,$$

Eq. 39

where $s_{ch}$ is the initial footpoint position (top of the chromosphere), $L$ is the initial chromosphere-to-apex half-length, $Q_{s,i}$ and $Q_{w,i}$ are the heating rates at $s_{ch}$ on the strongly and weakly heated sides, respectively, and $Q_{bkg}$ is a uniform background heating of much smaller magnitude. By convention, the strongly heated side has smaller $s$ and is on the left in our plots. It is the side with greater pressure as determined from Equation (24). The heating rate is uniform in the deep chromosphere, but the exponential variation extends to a depth of 5 Mm below $s_{ch}$ to allow for any downward displacement.

Although observationally distinct coronal loops seem to have a constant width (Klimchuk 2000; Lopez Fuentes, Klimchuk, & Demoulin 2006), most of the coronal plasma is contained in the diffuse component. The magnetic field must diverge with height, on average, and therefore the geometry we adopt for most of our runs has a cross-sectional area that expands linearly by a factor of 3 from 5 Mm below $s_{ch}$ to the apex. We smooth the area profile at the top, as shown in Figure 4.

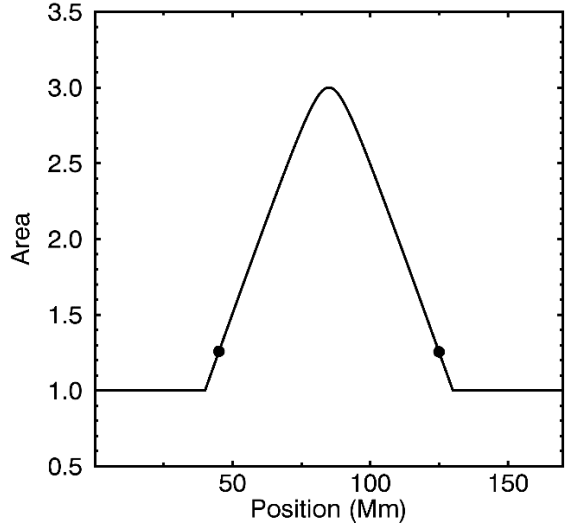

**Figure 4.** Cross sectional area as a function of position along the loop for simulations with minimal expansion in the low corona (a majority of our cases). The dots indicate the initial position of the top of the chromosphere.

The predictive formulae indicate that loop behavior is most affected by area expansion that occurs within the first heating scale length, so we consider several additional cases with strong expansion at low coronal altitudes. Such expansion is not unexpected (Guarrasi et al. 2014). To prevent downward displacement of the chromosphere from significantly affecting the expansion factor, which we wish to control, we have designed an area profile with minimal expansion below the initial footpoint position. The left side has the functional form

$$\frac{A(s)}{A_a} = \left(1 - \frac{A_{ch}}{A_a}\right) \frac{tanh\left(\frac{s - s_{ch} - w}{w}\right) + tanh(1)}{tanh\left(\frac{L - w}{w}\right) + tanh(1)} + \frac{A_{ch}}{A_a}$$

Eq. 40

and there is a similar form for the right side. Here, $A_a$ is the area at the apex, $A_{ch} = A(s_{ch})$ is the area at the initial footpoint position, and $w = 5$ Mm is the characteristic half-width of a steep gradient section in the low corona. Figure 5 shows an example where $A_a/A_{ch} = 5$ on the left side and 2.5 on the right.

The ARGOS code employs an adaptively refined mesh. Most of our simulations have approximately 3300 grid cells, with the smallest and largest cells being 4.1 and 66 km in size. Johnston et al. (2019) have shown that period of TNE cycles can be affected by an under-resolved transition region in which the minimum cell size is larger than about 2 km. A minimum cell size of several tens of km can even result in an artificial equilibrium when TNE is expected. We have repeated two of our simulations---our base model with TNE (model 1) and one of our steady flow models that is close to the TNE threshold (model 8)---and find that two additional grid refinement levels (1 km cells) do not lead to any significant change in behavior.

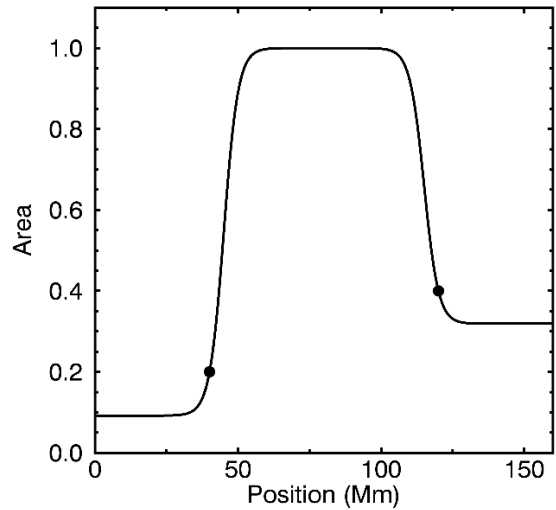

**Figure 5.** Cross sectional area as a function of position along the loop for a simulation with large expansion in the low corona. The dots indicate the initial position of the top of the chromosphere. In this example, the expansion factor is twice as large on the left as on the right.

### 4.1 Symmetric Loops

We begin by exploring loops with symmetric heating and symmetric area. The two possible behaviors are static equilibrium and TNE. Model 1 is our base model and is an example of TNE. It has the parameters: $Q = 0.037$ erg cm$^{-3}$ s$^{-1}$, $\lambda = 8.0$ Mm, $L = 42$ Mm, $Q_{bkg} = 0.0$ erg cm$^{-3}$ s$^{-1}$, $\Gamma = 1.2$ (linear expansion, Fig. 4). All other models have similar parameters except as noted. Only the first condition for TNE need be considered for symmetric loops. TNE is predicted if the ratio of the left hand side (LHS) to the right hand side (RHS) of Equation (36) is less than one. In this case the ratio is 0.066, so the prediction is correct.

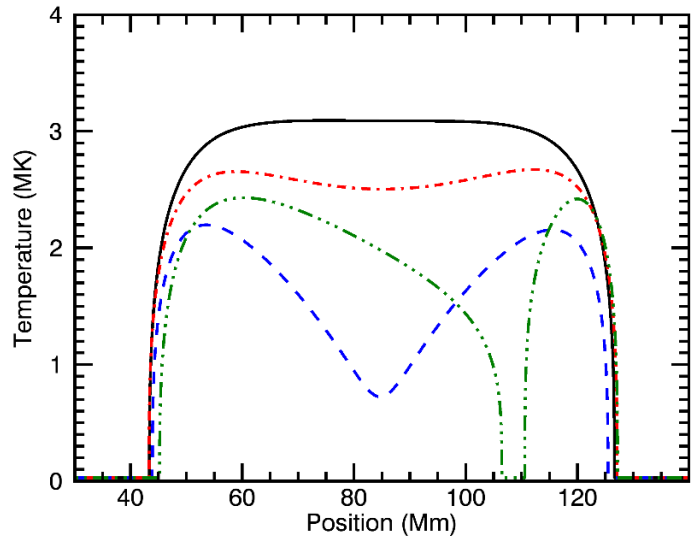

**Figure 6.** Temperature versus position along the loop at four times during model 1: 0 s (solid black), 1100 s (dot-dash red), 2400 (dash blue), and 12,600 s (triple dot-dash green) after $t = 175,500$ s from the start of the simulation. The loop is symmetric with $Q_{min}/Q = 0.011$. A movie version is also available, with time given relative to $t = 175,500$ s. The video begins at t=0 s, ends at t=24,400 s, and shows two TNE cycles. Its duration is 8 seconds of actual time.

27Figure 6 shows profiles of temperature versus position at four different times relative to $t = 175,500$ s (the transition from uniform to exponential heating happens at $t = 100,000$ s). This is a classic example of TNE behavior. Material evaporates into the loop as it attempts to establish an equilibrium in the strongly heated lower legs. This results in excess radiation in the upper part of the loop, and the plasma there slowly cools. A temperature dip forms at the apex, which rapidly accelerates from a thermal runaway, resulting in the creation of a cold, dense condensation. Symmetry is broken by numerical errors and/or low-amplitude waves in the system, and the condensation is nudged to the right (in other cycles it is nudged to the left). It slides down the leg to the footpoint, largely evacuating the corona in the process. The plasma then rapidly reheats due to the weakened radiation at the diminished densities, and a new cycle begins.

Table 1: Heating Concentration Formula, Equation (36)

| Model | Parameter Ratio | LHS / RHS | Actual Behavior |
|---|---|---|---|
| 1 | $Q_{min}/Q = 0.011$ | 0.066 | TNE |
| 2 | $Q_{min}/Q = 0.28$ | 1.9 | Static Equil. |
| 3 | $Q_{min}/Q = 0.28$, $\Gamma = 3.6$ | 0.95 | TNE |
| 4 | $Q_{min}/Q = 0.18$ | 1.1 | Static Equil. |
| 5 | $Q_{min}/Q = 0.023$ | 0.14 | TNE |

An examination of Equations (34) and (36) reveals that most of the parameters in our predictive formulae appear as ratios. We vary these ratios separately in order to isolate their effect on the loop behavior. The results for symmetric loops are listed in Table 1, with the base model in the first row. The critical ratio here is the minimum heating rate divided by the footpoint heating rate, $Q_{min}/Q = 0.011$. The four columns in the table give the model number, key parameter ratio, ratio of LHS to RHS in Equation (36), where ratios < 1 predict TNE, and the actual behavior. In all cases the parameter values are chosen to produce a peak loop temperature of roughly 2-3 MK.

Model 2 is similar to the base model, except that the heating scale length is increased by a factor of 2.5 at the same time that the footpoint heating rate is decreased in order to maintain a peak temperature in the desired range. Now $Q_{min}/Q = 0.28$. The heating stratification is too small to produce TNE, and static equilibrium is correctly predicted by the formula.

Model 3 is similar to model 2 except that the area expansion in the lower leg is increased to $\Gamma = 3.6$ (using the profile of Eq. [40]). This is enough to destroy the equilibrium and induce TNE, as predicted. Expansion makes symmetric loops more prone to TNE because it increases the density compared to uniform cross section.

Because the background heating is set to zero in these first three cases, a convenient measure of the heating stratification is the ratio of the heating scale length to the loop half-length, $\lambda/L$. It has values of 0.19 in model 1 and 0.51 in models 2 and 3. Winebarger et al. (2003) examined the dependence on this ratio for a large number of loops with exponential heating and uniform cross section. They found a critical ratio $\lambda/L = 0.45$ separating TNE from static equilibrium in loops having the same length and temperature as models 1 and 2, consistent with our results.

A uniform background heating is turned on in models 4 and 5, with everything else the same as in the base model. The additional background heating is enough to prevent TNE in model 4, but not in model 5, as correctly predicted. It appears that the critical ratio of apex-to-footpoint heating rate is somewhere near 0.1 in loops with small area expansion in the low corona. The importance of uniform background heating in preventing TNE has also been emphasized by Johnston et al. (2019).

## 4.2 Asymmetric Loops

We next examine the effect of asymmetries. Table 2 is similar to Table 1, except that the third column gives the ratio of the LHS to the RHS of Equation (34), using $\delta$ from Equation (37). The two possible behaviors are now TNE and steady flow. Static equilibrium is precluded by the asymmetries. The first condition for TNE, Equation (36), is satisfied for all models in the table.

Models 6, 7, and 8 are similar to the base model except that the footpoint heating rate is increased and decreased on the left and right sides, respectively, producing a $Q_s/Q_w$ asymmetry. As correctly predicted, the asymmetry is too small to prevent TNE in model 6, but it is large enough in model 7. Figure 7 shows the temperature profile at four times during model 6. Because of the pressure imbalance, the condensation forms off center and is always pushed to the right. The formation time is similar to the symmetric base model (~ 2500 s since the end of the previous cycle, when the condensation reaches the footpoint), but the time required to "fall" to the chromosphere is much shorter: 1300 s versus 10,000 s.

Table 2: Asymmetry Formula, Equation (34)

| Model | Parameter Ratio | LHS / RHS | Actual Behavior |
|---|---|---|---|
| 6 | $Q_s/Q_w = 1.5$ | 0.22 | TNE |
| 7 | $Q_s/Q_w = 7.4$ | 1.1 | Steady Flow |
| 8 | $Q_s/Q_w = 3.9$ | 0.55 | Steady Flow |
| 9 | $\lambda_s/\lambda_w = 2$ | 0.67 | TNE |
| 10 | $\lambda_s/\lambda_w = 6$ | 1.5 | Steady Flow |
| 11 | $\Gamma_s/\Gamma_w = 2.4$ | 0.69 | TNE |
| 12 | $Q_{min}/Q_s = 0.14, \Gamma_s/\Gamma_w = 2.1$ | 1.3 | Steady Flow |

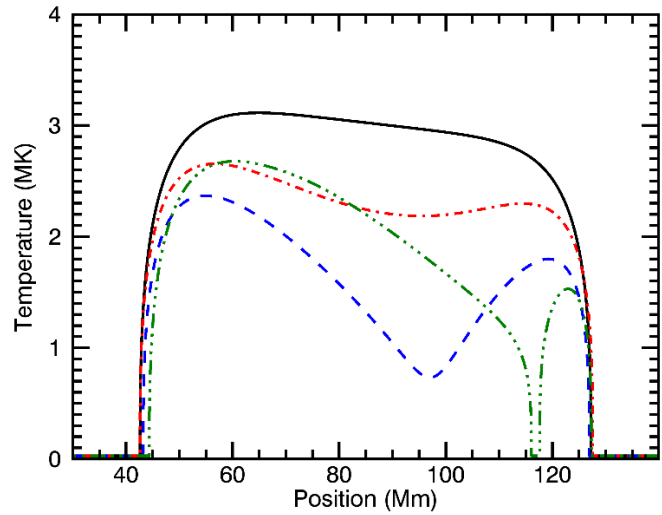

**Figure 7.** Temperature versus position along the loop at four times during model 6: 0 s (solid black), 1600 s (dot-dash red), 2600 (dash blue), and 3600 s (triple dot-dash green) after t = 183,800 s from the start of the simulation. The loop is asymmetric with $Q_s / Q_w = 1.5$. A movie version is also available, with time given relative to $t = 183,800$ s. The video starts at t=0 s, ends at t=16,000 s, and shows three TNE cycles. Its duration is 5 seconds of actual time.

Figure 8 shows the steady state temperature profile of model 7. The shape is similar to the sketch in Figure 2, with sharply rounded knees and a long flat section.

Model 8 has a heating asymmetry intermediate between models 6 and 7. TNE is predicted, but steady flow is observed. Thus, Equation (34) underestimates the ability of the footpoint heating asymmetry to prevent TNE in this case. Based on models 6 and 8, the critical ratio for $Q_s/Q_w$ is somewhere between 1.5 and 3.9 (when the other parameters are similar to the base model).

Models 9 and 10 are similar to the base model except that the heating scale length is shortened in the right leg to give a scale length ratio of $\lambda_s/\lambda_w = 2$ and 6, respectively. As correctly predicted, the larger ratio is enough to prevent TNE, but the smaller ratio is not. Figure 9 shows the temperature profile at four times during model 9. This example is somewhat atypical, though not rare. The condensations come in pairs---not two at the same time, but rather two occurring in rapid succession followed by a longer delay before the next pair. In addition, the temperature has a permanent dip, never fully recovering to become concave downward everywhere, is in most other cases. The dip and the condensations occur rather low in the leg, approximately one-third of the way to the apex from the right (weakly heated) footpoint. This is indicative of loops that are close to the critical condition separating TNE from steady flow. We also note that model 10 is not a perfect steady state. There is a small level of "wiggling" in the temperature profile.

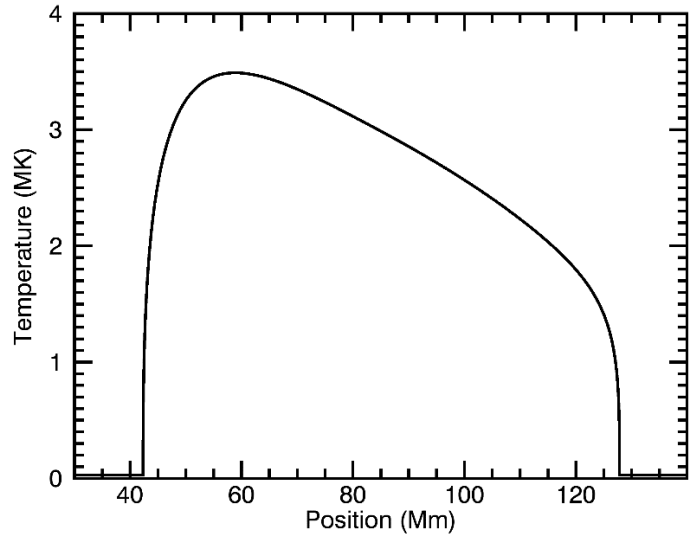

**Figure 8.** Temperature versus position along the loop in model 7. The loop is asymmetric with $Q_s / Q_w = 7.4$.

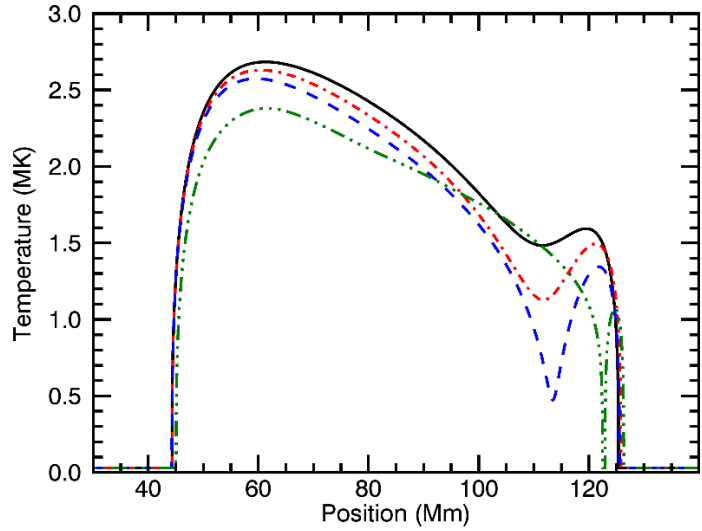

**Figure 9.** Temperature versus position along the loop at four times during model 9: 0 s (solid black), 600 s (dot-dash red), 900 (dash blue), and 1300 s (triple dot-dash green) after t = 188,000 s from the start of the simulation. The loop is asymmetric with $\lambda_s / \lambda_w = 2$. A movie version is also available, with time given relative to $t = 188,000$ s. The video starts at t=0 s, ends at t=12,000 s, and shows three pairs of TNE cycles. Its duration is 4 seconds of actual time.

Model 11 includes a $\Gamma_s/\Gamma_w = 2.4$ asymmetry in expansion factor that is not present in the base model. This is not enough to prevent TNE, as predicted. However, if the heating stratification is decreased in both legs compared to the base model, such that $\lambda/L = 0.38$ instead of 0.19 ($Q_{min}/Q = 0.14$ instead of 0.011), the conditions are close enough to equilibrium that $\Gamma_s/\Gamma_w = 2.1$ is adequate to drive a steady flow, as correctly predicted by model 12. Note that Equation (36) is satisfied, so the decrease in stratification would not by itself prevent TNE.

We have run a total of 30 models, most not shown here, with a variety of parameter combinations and peak coronal temperatures. 80% of them behave in the manner predicted by Equations (34) and (36). Most of the failed models do so by a reasonably small margin (e.g., model 8). We consider this a very satisfactory agreement given the approximations involved in the derivations.

## 5. Discussion

Using straightforward physical arguments and an innovative application of static equilibrium scaling laws, we have derived two formulae for predicting whether thermal non-equilibrium will occur. The conditions for TNE are: (1) the heating must be sufficiently concentrated at low coronal altitudes, and (2) asymmetries in the heating and/or cross-sectional area cannot be too great. As a rough rule of thumb, the ratio of minimum heating rate to footpoint heating must be less than about 0.1, and the asymmetry must be less than about a factor of 3. The precise values are case dependent, however.

We are pleased with the overall good agreement between the predictive formulae and our simulations, but it must be remembered that the simulations are quite idealized. Tests should be made with more realistic models having more complex variations of heating rate and cross-sectional area with position along the loop. Work along these lines was recently begun by C. Downs and C. Froment (private communication), and the initial results are encouraging. These more realistic simulations use a radiative loss function more complex than the single power law used in the formulae and our tests. Fortunately, we do not expect the details of the loss function to play a major role, since its slope, $b$, does not appear in Equation (13), and there is only a weak dependence in Equation (27). The coefficient $\Lambda_0$ is more important (Equation [28]).

We note that Froment et al. (2018) find, using realistic magnetic geometries and a realistic loss function, that TNE is more likely if the footpoint heating rate is increased with all other parameters held constant. Because the uniform background heating is fixed, increasing the footpoint heating decreases the ratio $Q_{min}/Q_s$. Both of our formulae predict that TNE is more likely in this situation, consistent with the simulations.

A number of other approximations and simplifications were used to derive the formulae that could be more significant. Chief among them are the constant $c_1$ and the coefficient $(2/3)^{1/2}$ in the expression for $\xi$ in Equation (32). Neither of these are strictly constant, but rather vary from loop to loop and as a function of time within a given loop. Recall that $c_1$ is the ratio of the integrated (along the loop) radiative losses in the transition region and coronal portions of the imaginary mini-loop used to estimate the conditions in the lower leg of the actual loop. Its behavior has been well studied in loops with uniform cross section (Klimchuk et al. 2008; Cargill et al. 2012), but not in expanding loops. C. Downs (private communication) suggests that defining $c_1$ without the area normalization, i.e., no $A_c$ in Equation (2), may improve the predictions. However, as pointed out earlier, comparison with the analytical solutions of Martens (2010) indicates that the area normalization is appropriate. Cargill et al. (2012) showed that $c_1$ depends on apex temperature when gravitational stratification is taken into account, but there should be minimal stratification in the imaginary mini-loops where $c_1$ is applied. Winebarger et al. (2003) report that the critical ratio $\lambda/L$ for TNE in symmetric loops has a weak dependence on apex temperature and loop length. There is no explicit dependence on either in Equation (13), suggesting that the dependence is incorporated into a non-constant $c_1$.

We have computed $c_1$ for the mini-loop segment (extending one heating scale length above the transition region) in our base model at a time when the temperature profile is flat (solid curve in Figure 6). This seems an appropriate time, since equilibrium loops with a flat temperature profile are at the threshold for TNE, and it is this threshold that our formulae seek to identify. The computed value of $c_1$ is 2.4, not far from the adopted 2.0.

The net influence of the approximations leading to $\xi$ are more difficult to evaluate. Examination of Equation (21) reveals that $\xi$ is the Mach number one heating scale length above the transition region in the strongly heated leg, defined in terms of the isothermal sound speed. Our steady flow simulations that are closest to TNE have a Mach number of ~ 0.15, whereas the value of $\xi$ predicted by Equation (32) is roughly 3 times larger. This difference is an indication of the errors in the approximations.

Although the logic underlying the asymmetry formula is straightforward, with end-to-end flow being driven by a simple pressure differential, the actual flow pattern that gets set up in the equilibrium is rather complex. Our strategy of using mini-loops assumes that the flow is energetically ignorable in the lower legs of the full loop, but this is not always the case. A modest Mach number means that kinetic energy is small compared to enthalpy, but it does not imply that the enthalpy term is negligible in the energy equation. The actual flows turn out to be fast enough that the temperature and especially the pressure given by Equations (14) and (15) are highly approximate. In particular, the pressure is more uniform along the loop than the difference between $P_s$ and $P_w$ in Equation (15) would imply. We have begun to consider an improved version of the asymmetry formula that is based on the energetics that drive evaporation on the strongly heated side, but further work is required. In any event, the success of our current formula suggests that it captures most of the essential physics.

Our 1D hydrodynamic treatment assumes that the magnetic flux tube confining the plasma is perfectly rigid, preventing waves and other external influences from affecting the behavior. Observations and MHD simulations suggest that flux tubes undergoing TNE cooling can induce sympathetic cooling in neighboring flux tubes (Antolin & Rouppe van der Voort 2012; Fang et al. 2015). The exact nature of the sympathetic cooling has not been established, but the closeness of the flux tubes to the TNE threshold could be important. We will comment on this further in our forthcoming paper.

We stress that are formulae are not expected to be accurate in tall and/or cool loops where pressure stratification is large. Nor are they likely to be accurate in flux tubes with extreme area variation in the upper legs, such as occurs near magnetic null points, for example.

The results we have presented here are important because TNE has been implicated in a number of interesting phenomena (prominences, coronal rain, long-period pulsating loops, and streamer tip pinch-off events). It has been suggested that it may be more widespread than these relatively isolated phenomena, perhaps even applying to a significant fraction of the corona. We have found that asymmetries are quite effective at preventing TNE, but the extent to which asymmetries of the required magnitude are present in the actual corona has not been carefully studied. We highly recommend that such an investigation be undertaken. Observational determination of the fraction of coronal magnetic flux undergoing TNE behavior would provide valuable information on the spatial and temporal properties of coronal heating. This would constrain the possible physical mechanisms of heating, shedding new light on this fundamental question.

Acknowledgements

We acknowledge useful discussions with many individuals, but especially members of International Space Science Institute team on "Observed Multi-Scale Variability of Coronal Loops as a Probe of Coronal Heating," led by C. Froment and P. Antolin. We thank Peter Cargill and the anonymous referee for useful comments. The work of J.A.K. was supported by the NASA Supporting Research program and GSFC Internal Scientist Funding Model (competitive work package) program. The work of M.L. was supported by the Spanish Ministry of Economy and Competitiveness (MINECO) through project AYA2014-55078-P and under the 2015 Severo Ochoa Program MINECO SEV-2015-0548. J.A.K. thanks the staff of Calverton Panera for creating a productive work environment.